\documentclass[12pt,a4paper]{article}
\usepackage{latexsym,amsmath,amssymb}
\date{}
\numberwithin{equation}{section}
\newtheorem{theorem}{Theorem}

\begin{document}
\title{Supersymmetric K\"allen-Lehmann representation}
\author{Florin Constantinescu\\ Fachbereich Mathematik \\ Johann Wolfgang Goethe-Universit\"at Frankfurt\\ Robert-Mayer-Strasse 10\\ D 60054
Frankfurt am Main, Germany}
\maketitle

\begin{abstract}
We find the general form of supersymmetric invariant two point functions. By imposing supersymmetric positivity we obtain the general supersymmetric K\"allen-Lehmann representation. 
\end{abstract}

\section{Introduction}

The K\"allen-Lehmann representation \cite {We} gives the general form of the two point functions in quantum field theory in terms of a positive spectral measure. It summarizes two of the most important ingredients of quantum field theory: Poincare invariance and positivity (combined with a certain amount of singularity which is indispensable in quantum field theory see \cite{Wi}). For a (free or interacting) complex scalar field $\Phi (x) $ it is
\begin{equation}
(\Phi (x)\Omega ,\Phi (y)\Omega )=(\Omega ,\Phi^+ (x)\Phi (y)\Omega)=\int_ {0}^{\infty }dm^2 \mu (m^2)\Delta _+(x-y ,m^2) 
\end{equation}
where  $\Delta _+ $ is the familiar Pauli-Jordan function (for notations and conventions in the non-supersymmetric case see \cite {We}) and $\mu (m^2) $ a positive measure of polynomial increase.
As noted above the K\"allen-Lehmann representation is a consequence of Poincare invariance and positive definitenes of the distribution $ (\Phi (x) \Omega ,\Phi (0)\Omega ) $. A straightforward proof follows  in the distribution-theoretic approach by using the general form of Lorentz invariant distributions \cite{BLOT}.
It is surprising that a similar representation for supersymmetric
theories doesn't appear in the literature. Here we want to fill this
gap by obtaining the  K\"allen-Lehmann representation for
supersymmetric (scalar) fields. The paper is structured as follows: in the second section we find the most general supersymmetric invariant two point functions, in the third section we discuss the supersymmetric positivity and in the forth section we obtain the K\"allen-Lehmann representation. We conclude the paper with a remark concerning the supersymmetric generalized free field and some other comments.

\section{Supersymmetric invariant two point functions}

Let $F(z)$ and $F(z_1,z_2)$ be supersymmetric functions (or distributions) of variables $z(x,\theta ,\bar \theta )$ (we use the notations and conventions of \cite {WB}; the notations in \cite {S} are the same as in \cite {WB} with the exception of the Pauli $\sigma _0$ which is minus identity in \cite {WB} and plus identity in \cite {S}).
Such functions can be expanded in powers of $\theta ,\bar \theta $. For example

\begin{gather} \nonumber
F(z)=F(x,\theta ,\bar \theta )= \\ \nonumber
=f(x)+\theta \varphi (x) +\bar \theta \bar \chi (x) +\theta ^2m(x)+\bar \theta^2n(x)+ \\
\theta \sigma^l\bar \theta v_l(x)+\theta^2\bar \theta \bar \lambda(x)+\bar \theta^2\theta \psi (x)+ \bar \theta^2 \theta^2d(x)
\end{gather}
A similar expression can be given for $F(z_1,z_2)$. Now suppose $F(z_1,z_2)$ is invariant under simultanous supersymmetric transformations of the variables $z_1,z_2$. \\
We want to find the two-point functions (i.e. functions $F(z_1,z_2)$
dependind on two variables $z_1=(x_1,\theta_1 ,\bar \theta_1
),z_2=(x_2,\theta_2 ,\bar \theta_2 )$) which are invariant under an
arbitrary simultaneous supersymmetric transformation of the
variables. Denoting by 
$Q=(Q_{\alpha }),\bar Q=(\bar  Q_{\dot \alpha })$ the generators of the supersymmetric transformations (up to a factor of $i$):
\begin{gather}
iQ_{\alpha}=\partial_{\alpha }-i\sigma_{\alpha \dot \alpha }^l 
\bar\theta^{\dot \alpha }\partial_l \\
i\bar Q_{\dot \alpha }=-\bar \partial_{\dot \alpha }
+i\theta ^{\alpha }\sigma_{\alpha \dot \alpha }^l \partial_l 
\end{gather}
simultaneous supersymmetric invariance imply

\begin{gather}
(Q_1+Q_2)F(z_1,z_2)=0 \\
(\bar Q_1+\bar Q_2)F(z_1,z_2)=0
\end{gather}
where $Q_1,Q_2$ and $\bar Q_1,\bar Q_2$ act on the variable $z_1$ and $z_2$ respectively.\\
In order to solve these equations we introduce new variables $\theta=\frac {1}{2}(\theta _1+\theta _2) $
and $\zeta =\theta _1-\theta _2 $ together with their conjugates as well as (by translation invariance) the difference variable $x=x_1-x_2 $ (a somewhat similar argument but in a different context also appears in \cite{GS}). Note that by introducing the difference variable $x=x_1-x_2 $ the derivative $\partial _l $ in $Q_2,\bar Q_2 $ taken with respect to the second variable changes sign such that in the new variables equations (2.4),(2.5) take the form

\begin{gather}
(\frac {\partial }{\partial \theta ^{\alpha }}-i\sigma_{\alpha \dot \alpha }^l \bar \zeta^{\dot \alpha }\partial_l )F=0   \\
(\frac {\partial }{\partial \bar \theta^{\dot \alpha }}-i\zeta^\alpha \sigma_{\alpha \dot \alpha }^l \partial_l )F=0 
\end{gather}
where  F is a function depending on the variables $x=x_1-x_2 ,\theta ,\bar \theta , \zeta ,\bar \zeta $ . We want to solve this system of partial differential equations in mixed commutative and non-commutative variables.
A first (trivial and from the physical point of view uninteresting)
solution for $F$ is a constant. Other solutions can be obtained in a
two step procedure by using in the first step the equation (2.6) to factorize from $F$ the exponential $ exp (i\theta \sigma^l\bar \zeta \partial_l )$. We write

\begin{equation}\nonumber
F=exp (i\theta \sigma^l\bar \zeta \partial_l )D
\end{equation}
The first equation (2.6) implies $ \frac {\partial }{\partial \theta ^{\alpha }}D=0 ,\alpha =1,2 $ which means that $ D $ is independent of $ \theta_{\alpha },\alpha =1,2 $. In the second step we write

\begin{equation}\nonumber
D=exp (-i\zeta \sigma^l \bar \theta \partial_l )E
\end{equation}
and conclude as above that $E$ is independent not only of $\theta $ but also independent of $ \bar \theta $.
These two exponentials cover the dependence of $F$ from the variables
$\theta $ and $\bar \theta $. The residual dependence in $E$ is in
$\zeta ,\bar \zeta $ and $x=x_1-x_2$. Altogether the general solution
of the equations (2,6),(2.7) in the $x,\theta ,\bar \theta ,\zeta ,\bar \zeta $-variables is (with the exception of the constant solution) of the form 

\begin{equation}
F(x,\theta ,\bar \theta ,\zeta ,\bar \zeta )=exp[\pm i(\zeta \sigma^l\bar \theta -\theta \sigma^l\bar \zeta )\partial_l ]E(x,\zeta ,\bar \zeta )
\end{equation}
where from invariance considerations $E$ is restricted to

\begin{equation}
E(x,\zeta ,\bar \zeta )=E_1(x)+\zeta^2E_2(x)+\bar \zeta^2E_3(x)+\zeta \sigma^l\bar \zeta \partial_lE_4(x)+\zeta^2 \bar \zeta^2 E_5(x)
\end{equation}
with $E_i(x)=E_i(x_1-x_2),i=1 \ldots 5 $  Lorentz invariant functions
(or distributions). We obtain in this way (with the exception of the
trivial constant) a total of 5 linearly independent (in the Grassmann
variables)  invariants. Note the $\pm $ sign in front of the
exponential in (2.8). The supersymmetric invariance in (2.4),(2.5) is
formulated in terms of the left multiplication. It produces the minus
sign in (2.8). We could have used right multiplication in which case
the equations (2.4),(2.5) are replaced by

\begin{gather}\nonumber
(D_1+D_2)F(z_1,z_2)=0 \\ \nonumber
(\bar D_1+\bar D_2)F(z_1,z_2)=0 \nonumber
\end{gather}
where $D,\bar{D} $ are the covariant derivatives \cite{WB}. By solving
these equations in the same manner as the equation (2.6),(2.7) we produce the plus sign in the exponential in (2.8).\\
Although the problem seems to be solved it is clear that we didn't obtain yet the result we wish to obtain. Indeed there are a bunch of invariants easy to construct by hand and which are intensively used in physics. We will give them below. We want to recognize them among the solutions of the partial differential equations above and to investigate if there are some others invariants too. We start with the "supersymmetric Pauli-Jordan" function
\[ K_0=\delta (\theta_1-\theta_2 )\delta (\bar \theta_1-\bar \theta_2 )\Delta_+ (x_1-x_2) \]
It is supersymmetric invariant and it can be located among the invariants deduced as solutions of the partial differential equations as the one containing the $\delta- $ function product $\zeta^2 \bar \zeta^2 $. Certainly the exponential in (2.8) disappears because of the $\delta -$function. In order to see directly that $K_0$ is supersymmetric invariat we can apply on both variables $z_1,z_2$ the supersymmetric transformations
\begin{gather}
x^{'m}=x^m+i\theta \sigma^m \bar \epsilon -i\epsilon \sigma^m \bar \theta \\
\theta^{'}=\theta +\epsilon , \bar \theta^{'}=\bar \theta +\bar \epsilon
\end{gather}
with increments $\epsilon_1 ,\epsilon_2 $.
The invariance follows after disregarding third powers of $\theta
_1-\theta _2 $ and $\bar \theta_1-\bar \theta_2 $ variables. This means that the even Grassmann part in (2.10) has no influence on $\Delta_+ (x_1-x_2) $ which immediately implies the invariance of $K_0$.  \\
We continue by giving other invariants used in physics which can be constructed from $K_0$. For simplifying the notations in the rest of the paper let us denote 

\begin{equation}
\Delta=\Delta (\theta_1 -\theta_2 ,\bar \theta_1 -\bar \theta_2)=\delta (\theta_1 -\theta_2 )\delta (\bar \theta_1 -\bar \theta_2 )=(\theta_1 -\theta_2 )^2(\bar \theta_1 -\bar \theta_2 )^2 
\end{equation}
Fist of all note that by replacing in the supersymmetric Pauli-Jordan
function $K_0$ the function $\Delta_+$ by an arbitrary Poincare
invariant funtion (or distribution) $G_0$ we obtain other
supersymmetric invariants. All of them belong to the same family whose
elements will be denoted by $\Delta G_0=P_0\Delta G_0$ where for
notational reasons to be seen later we define $P_0=1$. Beside $K_0$
other five (families of) invariants can be obtained from $\Delta G_0$
(with different $G_0$) by applying to it the covariant and invariant \cite{S} 
derivatives $D^2,\bar D^2,D^2\bar D^2,\bar D^2D^2 $ and $ D^{\alpha
}\bar D^2D_{\alpha}$. Because $\Delta G_0 $ depends on the variables
$x_1,\theta_1,\bar \theta_1 $ and $x_2,\theta_2,\bar \theta_2 $ we
have to specify on which variables we apply the covariant derivatives.
It turns out that this is a minor problem. We will let them act on the
first variables $x_1,\theta_1,\bar \theta_1$ and take the plus sign in
the exponential (2.8). Other choises give no new results.\\
Whereas as discussed above it is relatively easy to see that $\Delta G_0$ is of the form (2.8),(2.9) it is not totally obvious that the members of the five derived families above are of the form (2.8),(2.9) First note that a compact way to write them down is by using the formal projection operators 

\begin{gather}
P_c=\frac {1}{16}\frac {D^2 \bar D^2}{\square } \\
P_a=\frac {1}{16}\frac {\bar D^2 D^2}{\square }\\
P_T=-\frac {1}{8 \square }D^{\alpha }\bar D^2D_{\alpha}=-\frac {1}{8 \square }\bar D_{\dot \alpha }D^2\bar D^{\dot \alpha}
\end{gather}
and two more operators

\begin{gather}
P_+=\frac {D^2}{4\square ^{\frac {1}{2}}} \\
P_-=\frac {\bar D^2}{4\square ^{\frac {1}{2}}}
\end{gather} 
We do not comment on some technical aspects related to the
d'alembertian in the denominator. This problem has been discussed in detail in \cite{C1}. \\
The operators $P_i,i=c,a,T,+,-$ form a closed algebra under multiplication.\\
We recognize the invariants which we have constructed by hand to be of the form $P_i\Delta G_i,i=0,c,a,T,+,-$ where $G_i$ are Poincare invariants. The operators $P_1,P_2,P_3$ satisfy the relations \cite{WB}

\begin{gather}
P_c+P_a+P_T=1 \\
(P_c+P_a-P_T)\delta (\theta_1-\theta_2)\delta (\bar \theta_1-\bar \theta_2)= \\ \nonumber
=\frac {4}{\square }exp [i(\theta _2\sigma^l \bar \theta_1-\theta_1
\sigma^l\bar \theta_2)\partial_l ]
\end{gather}
Now we come back to (2.8),(2.9). By promoting the term $\zeta
\sigma^l\bar \theta \partial_l$ in (2.9) to the exponential and
adjusting in the linear hull of $E_1,\zeta^2 E_2,\bar \zeta^2
E_3,\zeta^2 \bar \zeta^2 E_5 $ we finally obtain the following result: \\

\begin{theorem}
The two-point supersymmetric invariant functions (distributions) are linear combinations of $P_i\delta (\theta_1-\theta_2)\delta (\bar \theta_1-\bar \theta_2 )G_i(x_1-x_2)$,where $G_i(x)$ ,$i=c,a,T,+,-$ are Lorentz-invariant functions (distributions).
\end{theorem} 
In order to prove this assertion we need the following identities \cite{WB}

\begin{gather}\nonumber
D_1^2(\theta_1 -\theta_2 )^2=-4exp[-i(\theta_1 -\theta_2 )\sigma^l \bar \theta_1  \partial_l^1 ]  \\ \nonumber
\bar D_1^2 (\bar \theta_1 -\bar \theta_2 )^2=-4exp[i\theta_1 \sigma^l (\bar \theta_1 -\bar \theta_2 ) \partial_l^1 ]  \\ \nonumber 
\bar D_1^2D_1^2(\theta_1 -\theta_2 )^2(\bar \theta_1 -\bar \theta_2 )^2=16exp[i(\theta_1 \sigma^l \bar \theta_1 +\theta_2 \sigma^l \bar \theta_2 -2\theta_1 \sigma^l \bar \theta_2 )\partial_l^1 ]  \\ \nonumber
D_1^2\bar D_1^2(\theta_1 -\theta_2 )^2(\bar \theta_1 -\bar \theta_2 )^2=16exp[-i(\theta_1 \sigma^l \bar \theta_1 +\theta_2 \sigma^l \bar \theta_2 -2\theta_2 \sigma^l \bar \theta_1 )\partial_l^1 ] \\ \nonumber
\end{gather}
where for the convenience of the reader we provided the derivatives with the index of the acted on variable (the first one).
These relations exactely provide the connection between the invariants
obtained by solving the partial differential equations (2.6),(2.7) and
the invariants given with the help of the covariant derivatives. In
particular from the $\zeta^2 ,\bar \zeta^2 -$terms in (2.9) we obtain
the $ P_{\pm }-$invariants and from the rest of the terms in (2.9) we obtain the $P_i,i=c,a,T-$invariants.\\
In other words we have rigorously proved that the expressions $P_i\Delta G_i$ are supersymmetric invariants (a fact which was clear from the very construction of these expressions) and, more important, that there are no other two-point supersymmetric invariants. Stated in a clear way the statement shows that besides the supersymmetric invariant two-point functions which already appear in the physical literature (and are easy to construct by hand) there are no others.

\section{Supersymmetric positivity}

Now after we obtained the invariant two point functions $F(z_1,z_2)$ we have to ask the important question which one of them is positive definite in superspace i.e. for which kernels $F(z_1,z_2)$ the (hermitean) sesquilinear form

\begin{equation}
\int d^8z_1d^8z_2\overline{f(z_1)}F(z_1,z_2)g(z_2)
\end{equation}
is positive for $f=g$. The bar includes besides the usual complex conjugation also the Grassmann conjugation. Here $d^8z=d^4xd^2\theta d^2\bar \theta  $.\\
First of all let us remark that for $F=F(z_1,z_2 )=P_{\pm
}\Delta(\theta_1 -\theta_2 ,\bar \theta_1 -\bar \theta_2 )G_{\pm } $
the integral (3.1) with $f=g$ does not have definite sign even if
$G_{\pm }$ is a positive definite distribution. This can be proven by
decomposing $f$ with the help of the projection identity
$P_c+P_a+P_T=1$ and usind the properties of the covariant derivatives
$D, \bar D$. It means that from the five invariants associated to
$P_i,i=c,a,T,+,-$ only three of them (those associated to
$P_i,i=c,a,T$) have to be taken into consideration. It was shown in \cite{C1}
that all of them can give rise to positive definite sesquilinear forms
(scalar products). Indeed $P_i\Delta G_i(z_1,z_2),i=c,a$ as well as
$-P_T\Delta G_T$ are positive definite supersymmetric kernels if
$G_i(x_1-x_2)$ are positive definite in the usual sense. Note the
minus sign in front of $P_T$ which seems to be at odd with the
projection identity $P_c+P_a+P_T=1$ and plays a considerable role
because it induces a natural, inherent Krein structure of
supersymmetries. The proof of these assertions \cite{C1} is mainly by
computation using the positivity of $-p^2$ and $\sigma p$ for the
momentum $p$ in the forward light cone. The interested reader can find more material on this inherent structure of the $N=1$ superspace and applications to supersymmetric quantum field theory without path integral methods, in particular canonical quantizatin in \cite{C1}. \\ 
We have obtained our second result:

\begin{theorem}
The two point supersymmetric positive invariant functions (distributions) are precisely the elements of the positive cone generated by $P_c\Delta G_c $, $ P_a\Delta G_a $ and $ -P_T\Delta G_T $.
\end{theorem} 
We are ready now to present in the next section the K\"allen-Lehmann representation for free or interacting supersymmetric quantum field theory.  
 
\section{K\"allen-Lehmann representation} 

Let $\Phi(z)$ be a scalar (possible complex) supersymmetric quantum field.
It is understood that it acts in a Hilbert space of supersymmetric states with a vacuum $\Omega $. In order to cope with the singularities of $\Phi(z)$ in $x$-space we perform the smearing with a supersymmetric function $\varphi (z)$ obtaining

\begin{equation}
\Phi (\varphi )=\int d^8z\Phi (z)\varphi (z)
\end{equation}
Denoting by $(.,.)$ the Hilbert space scalar product the two-point function of the quantum field $\Phi (z)$ is defined to be

\[ (\Omega ,\Phi^+ (z_1)\Phi (z_2)\Omega ) \]
where $\Phi^+ $ is the Hilbert space adjoint of $\Phi $. It follows that, as in the usual quantum field theory too, the two-point function $(\Omega ,\Phi^+ (z_1)\Phi (z_2)\Omega )$ has to be positive definite (in the distribution theoretic sense). By making contact to the result in section 3 we obtain

\begin{theorem}[K\"allen-Lehmann supersymmetric representation] 
The two point function (distribution) in a supersymmetric quantum field theory has the form

\begin{equation}
(\Omega ,\Phi^+ (z_1)\Phi (z_2)\Omega )=\lambda_c P_c\Delta G_c+\lambda_a P_a\Delta G_a-\lambda_T P_T\Delta G_T
\end{equation}
where $\lambda_i$ are positive constants and $G_i  $ have (spectral) representations of the form 

\begin{equation}
G_i=G_i(x_1-x_2)=\int_0 ^\infty d\mu_i (m^2)\Delta_+ (x_1-x_2)
\end{equation}
for $i=c,a,T $. If the spectral measures coincide then the representation reads 

\begin{equation}
(\Omega ,\Phi^+ (z_1)\Phi (z_2)\Omega )=(\lambda_c P_c +\lambda_a P_a-\lambda_T P_T)\Delta G
\end{equation}
where
\begin{equation}
G=G(x_1-x_2)=\int_0 ^\infty d\mu (m^2)\Delta_+ (x_1-x_2)
\end{equation}
\end{theorem}
We remind the reader that $\Delta=\delta (\theta_1 -\theta_2 )\delta
(\bar \theta_1 -\bar \theta_2 )$and $\Delta_+ =\Delta_+(x_1-x_2)$ is
the faminiar Pauli-Jordan function. \\
Note that the above result is valied for real or complex scalar
fields (an example is the free supersymmetric (real) massive vector
field in the unitary gauge; for the arbitrary gauge fixing term there
is still a representation of type (4.2) but because the lack of positivity, the constants $\lambda_i $ are not all of them positive \cite{C1}). If a quantum field theory has several fields (like for instance the Wess-Zumino model which involves chiral as well as antichiral fields, or fields with a Lorentz spin) the result may be different. For instance the positive definite kernel which appear in the K\"allen-Lehmann representation of the Wess-Zumino model is a two by two matrix kernel with nonvanishing entries (see \cite{C1}).

\section{Conclusions}

We obtained the most general form of supersymmetric invariants in two variables. Combining this result with supersymmetric positivity we studied the supersymmetric K\"allen-Lehmann representation.\\
Starting with our results one can construct the generalized supersymmetric field by applying the same construction as usual \cite{J}. \\
Note that the spectral measures in (4.9) may be different. Physical considerations may impose a unique spectral measure (4.5). In this case the spectral measures which are induced at the level of the multiplet components (for instance those of the spinor field components) will be related to each other. This may have some consequences on renormalization phenomena.\\
Acknowledgement: we thank M. Schork for discussions.


\begin{thebibliography} {99}

\bibitem {We} S. Weinberg, The Quantum Theory of Fields, vol II, Cambridge University Press, 1996

\bibitem {Wi} A.S. Wightman, Ann. Inst. Henri Poincare, Ser. A 1(1964),403

\bibitem {BLOT} N.N. Bogoliubov, A.A. Logunov, A.I. Oksak, I.T. Todorov, General Principles of Quantum Field Theory, Kluwer, 1990

\bibitem {WB} J. Wess, J.Bager, Supersymmetry and Supergravity, 2nd edition, Princeton University Press, 1992

\bibitem {S} P. Srivastava, Supersymmetry, Superfields and Supergravity: An Introduction, IOP Publishing, Adam Hilger, Bristol, 1986

\bibitem {GS} D.R. Grigore, G. Scharf, hep-th/0212026

\bibitem {C1} F. Constantinescu, Lett.Math.Phys, 62(2002),111 ;J.Phys.A:Math.Gen., 38(2005),1385 ;Mod.Phys.Lett.A, 20(2005),1239, hep-th/0505191  

\bibitem {J} R. Jost, The General Theory of Quantized Fields, American Mathematical Society, 1965













\end{thebibliography}
\end{document}